# DYNAMIC RESONANCE EFFECTS IN THE STATISTICAL DISTRIBUTIONS OF ASTEROIDS AND COMETS

© 2009 B.R. Mushailov, V.S. Teplitskaya

Sternberg State Astronomical Institute

Lomonosov Moscow State University

*Some principles in the distribution of Centaurs and the "Scattered Disk" objects, as well as the Kuiper belt objects for its semi-major axes, eccentricities and inclinations of the orbits have been investigated. It has been established that more than a half from them move on the resonant orbits and that is what has been predicted earlier. The divergence of the maximum in the observable distribution of the objects of the Kuiper belt for the semi-major axes with an exact orbital resonance has been interpreted.*

## Introduction

The existence of the Kuiper belt has been predicted by Edgeworth (1949) and Kuiper (1951). The Kuiper belt (or "the Kuiper bank") located at 30-50 a.u. away from the Sun.

The possibility of the existence of more than one "Kuiper belts" was predicted in [1, 2]. The location of a few beyond- Neptune belts is probable in the range of $100 \leq a < 1000$ a.u. from the Sun. Large bodies (around 1000 kilometers in diameter) outside the orbit of Neptune, which are being discovered nowadays, suggest the presence of a planets beyond the orbit of Neptune that haven't been discovered yet. In particular, there are areas predicted in work [3], where hypothetical major planets beyond the orbit of Neptune could be located, and one can also find there all the calculations for the areas of mean motions connected by the resonances with these hypothetical planets.

The significant number of works is devoted to research of centaurs and transneptunian objects. In a number of works allocate two basic dynamic classes: centaurs and scattered disk objects, and Kuiper belt objects. Last conditionally divided into the classical belt objects Kuiper and several subclasses of resonant objects [4]. It means a resonant objects with librational type of motion [5, 6]. The evolution of the orbits of these objects and modern building beyond- Neptune zone are assumed due to the dynamic evolution of Neptune [7, 8, 9].

However, in the majority of the works, works devoted to the study of evolution beyond- Neptune objects, integration is spent on time intervals (millions years) on which effects of dynamic instability are essentially shown, and therefore reliability of the results received within the limits of gravitational model, raises the doubts. Research within the limits of the concept of partial determinacy of dynamic evolution of objects beyond- Neptune belts based on the elliptic restricted three-body problem, taking into account the gravitational perturbations of Uranus, Saturn and Jupiter [10, 11, 12] allows to construct a reliable model based on analytical methods to predict the evolution of orbital parameters of objects beyond- Neptune belts, and indicate the most likely area to find new objects. Based on the results of [10], authors of work [13] identified a number of objects in those zones. In view of continuous growth of population found beyond- Neptune objects, and now remains valid ordering observed distributions of these objects on the orbital parameters. . Irrespective of, what character of movement (libration or circulation) resonant effects are caused by a commensurability of average movements gravitating objects. Consequently, contrary to a number of works, a part of so-called classical objects of Kuiper belt actually concern to resonant as move in orbital commensurabilities with major planets of Solar system. Despite of significant number of the works, the statistical distributions of objects of Kuiper belt [4-9] devoted to research, till now it was not given due attention to revealing and a theoretical substantiation of true laws in observable distributions beyond-Neptune objects and centaurs to that the present work is devoted.

**Resonances**

The condition of a two-frequency resonance means implementation of rational quasi-commensurability of frequencies of a kind: *[(k+l)n –kn']≤O($\sqrt{\mu}$)*, where *l*- is the resonance order, *k*- is the resonance multiplicity; *l, k*- are the natural numbers; *n', n*- are the frequencies (or mean motions) of passively gravitating and disturbing bodies, *µ*- is reduced mass. The resonant zone having the order of *2n'$\sqrt{\mu}$/(k+l)*, is localized about a point of the exact commensurability: *n/ n' = k / (k + l)* in case of *n <n'*, or *n/ n' = (k + l) / k* in case of *n> n'*. The commensurabilities of the first order (*l*=1; the case of the Linbland resonances) are characterized by the maximal resonant effect. Maximal amplitude of the resonant effect at fixed *l* is obtained at *k*=1. Orbital resonances connect movements of some major planets, satellites, asteroids, comets, meteoric streams of Solar system, extra solar systems.

Dynamic evolution of the orbits of many of these bodies can be correctly interpreted on the basis of resonant version of the classic three-body problem, taking into account the secular perturbations from external bodies.

As is evident from the work [11], analytic expression for the semi-major axis (*a*) of orbit of a passively gravitating body (a beyond-Neptune object) at the account of the secular perturbations from the giant planets Jupiter, Saturn and Uranus *(P$_i$=1,2,3)* analytic expression for the semi-major axis (*a*) orbit passively gravitating body (beyond- Neptune object) has the form:

$$\sqrt{a} = \sqrt{a_0} + \Delta , \qquad (1)$$

where *a$_0$* is a semi- major axis under the absence of disturbances of the giant planets.

$$\Delta = \Delta_0 \sum_{i=1}^{3} \frac{m_i}{a_i} \beta_i [L_{1/2}^{(0)}(\beta_i) + \beta_i D L_{1/2}^{(0)}(\beta_i)], \quad \Delta_0 = \frac{1}{3} \left( \frac{k}{(k+l)(1+\mu_\psi)} \right)^{2/3}, \beta_i = a_i / \gamma, \qquad (2)$$

*γ* – is the integrated constant, *µ$_\psi$* -is the mass of Neptune, and $m_{\overline{1,3}}$ and $a_{\overline{1,3}}$, mass and orbital semi-major axis of Jupiter, Saturn and Uranus recpectively. For mass adoption of solar masses and semi-major axis of orbit of Neptune accepted for unit, $L_{1/2}^{(0)}$ - the relevant coefficient Laplace.

From (1) it is clear, that the secular disturbances from the giant planets lead to the expression for the semi-major axis of an object which is correspondent to an exact orbital resonance.

$$a = \alpha_\psi * \left[ \left( \frac{k+l}{k} \right)^{1/3} - \Delta \right]^2 , \quad a_\psi \text{ is the semi-major axis of Neptune.} \qquad (3)$$

In case of the first order resonance (*l* = 1), the following variables are introduced: *x=$\sqrt{2\xi}$ cos(η), y =$\sqrt{2\xi}$ sin(η)*, values of *ξ* and *η* are defined in [11] and a solution of the elliptic restricted three-body problem, taking into account the secular perturbations from of giant planets is given to the integration of an autonomous canonical system of equations with one degree of freedom of the form: $\frac{dx}{d\tau} = \frac{\partial F}{\partial y}, \frac{dy}{d\tau} = -\frac{\partial F}{\partial x}$, with Hamiltonian *F = (x² + y²) ² + A (x² + y²) + Bx*. (4)

Where *A=[4/(k+1)]{$\sqrt{\gamma} - E^{-2} + \Delta$}, $E = \left[ \frac{k}{k+1} \sqrt{1+\mu_\psi} \right]^{1/6}$,*

$$\gamma = a^{-1} \left[ -k + (1+k)\sqrt{1-e^2} \cos i \right]^{-2}, I^2 = 4\sqrt{a}\sqrt{1-e^2} \sin^2(i/2), \qquad (5)$$

*I* and *γ* are the integrals of the problem; *a, e, i* are the semi-major axis, eccentricity and the inclination of the orbit of a beyond-Neptune object respectively. The expression for *B* is also defined in [11]. Stationary solutions for variables *x* and *y*, are defined by a system of algebraic equations:

*4x(x²+y²)+2Ax+B=0,*
*4y(x²+y²)+2Ay=0.* (6)

Thus the configuration of a phase plane is presented in figure 1.

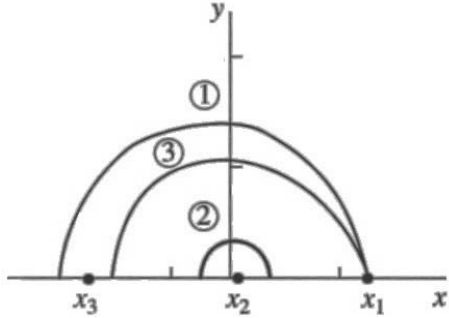

Fig. 1. Configuration of the phase plane. Trajectories passing through a fixed point ($x_1$, 0) are the separatrices, distinguish three characteristic area. Because of symmetry only the upper half is shown.

Being based on integrals of a problem (5), it is possible to present to area of instability on the diagram (*e,a*) and to compare with them to observed data (Fig.2).

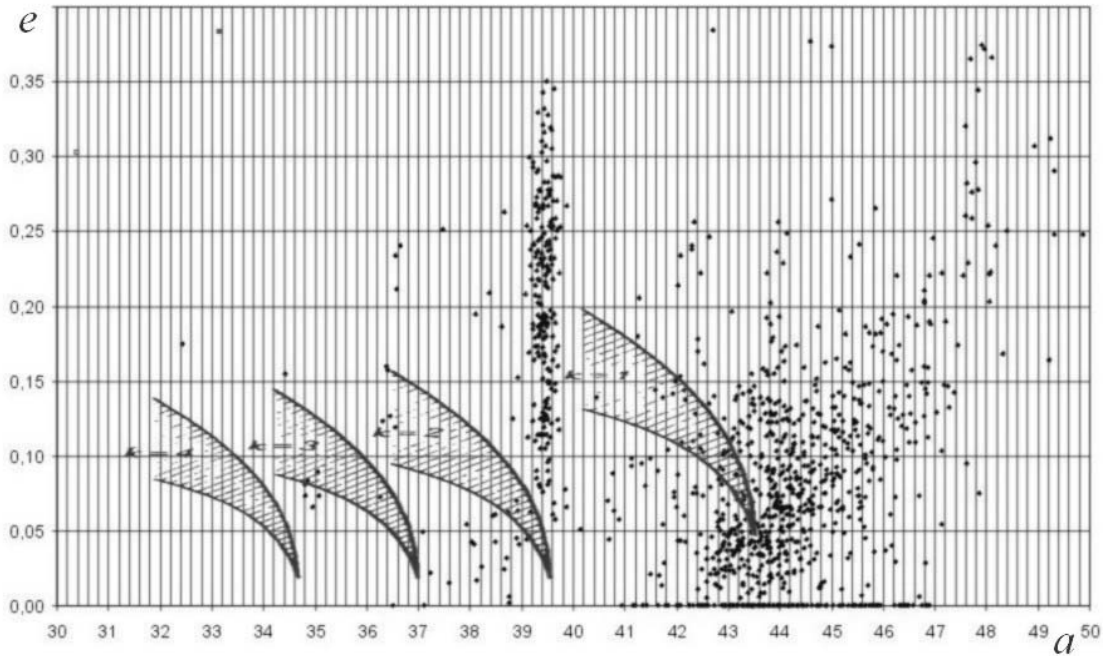

Fig. 2. Zones of instability in the diagram (*a, e*) - in the case of the first order resonances for the different multiplicities *k*.

If the elements of their orbits are located outside of the shaded areas, beyond-Neptune objects, concentrating at the borders of these zones, can exist for long time. Taking into account the non-symmetry of the zones of unsteadiness relative to values of the exact commensurability, it is easy to understand, that in resonant zones (ceteris paribus) the existence of beyond-Neptune objects (and with great values of eccentricity) is more probable, at $n < n_0 (k)$, than at $n > n_0 (k)$.

From fig. 2 follows, that found out beyond-Neptune objects are located in the plane of orbital parameters (*a,e*), according to theoretical results, outside areas of instability, except for 3.7 % of objects for a case of a resonance 2:1 and 0.55 % for a resonance 3:2 with the Neptune, which are most subject to migration.

For objects being orbital commensurabilities (4:3, 3:2, 2:1) with the Neptune on the basis of the orbital parameters given in [14], positions of the relevant image points in the phase plane were calculated. . It has appeared, that for a commensurability 4:3 in the libration zone (zone 3) there are 90 % of all found out objects, for a commensurability 3:2 - 83 % of objects, for a resonance 2:1, taking into account the secular perturbations of - 65% of objects.

Probabilities of transition into the libration zone can be assessed by exploring probabilities of transitions of trajectories from one areas of a phase plane in others under influence of various perturbing factors, characterized by independent parameters $\delta=(\delta_1 \delta_2 ... \delta_n)$. The probability of transition of a trajectory, under action of indignations, from area *i* in area *j* a phase plane is determined by expression [11]:

$$W_{ij} = \begin{pmatrix} (-1)^j \dfrac{2}{4-j} \Delta' + (j-3)\dfrac{\pi}{2} \\ (-1)^i \dfrac{\pi}{2} - \Delta' \end{pmatrix}, \quad i=1,2, \; j=2,3, \text{ were } \Delta' = \arcsin \varepsilon_1 + \psi_1 \varepsilon_2,$$

$$\varepsilon_1 = \dfrac{1}{2}\left(\dfrac{\psi_2}{1+sign(1-\psi_2)\varepsilon_3}\right)^{1/4}, \quad \varepsilon_2 = \dfrac{1}{\sqrt{3}}(1-4\varepsilon_3^2)^{1/2}, \quad \varepsilon_3 = \cos\left(\dfrac{\pi + \arccos|1-\psi_2|}{3}\right).$$

Here $\psi_1$ and $\psi_2$ - independent parameters related to the coefficients of the Hamiltonian (4) as follows: $\psi_1 = -\dfrac{A}{B}\dfrac{\partial B/\partial \delta}{\partial A/\partial \delta}$, $\psi_2 = -\dfrac{27}{4}\dfrac{B^2}{A^3}$.

Factor $B$ (at fixed $\mu_\psi$) depends on the multiplicity and the order of resonance, and $A$ - function of a condition of the system, characterized by integrated constants.

Probabilities of transition beyond-Neptune object of the zone i in zone j, depending on ctg$\Delta$ ($\delta$) are presented in Figure 3. For investigated objects of probability of transitions from one areas of phase space in others essentially depend on parameter $\rho=(\partial B/\partial \delta)/(\partial A/\partial \delta)$. For example, the probability of transition to a third area, for the resonances 4:3 and 3:2: $W_{13} > W_{12}$, with significance $\rho$ from -0.05 to -0.03, and $W_{23} > W_{21}$ - from 0.03 to 0.05. In the case of resonance 2:1 $W_{23} > W_{21}$, with $\rho$ of -0.018 to 0.011 and $W_{13} > W_{12}$, with $\rho$ from 0.01 to 0.02.

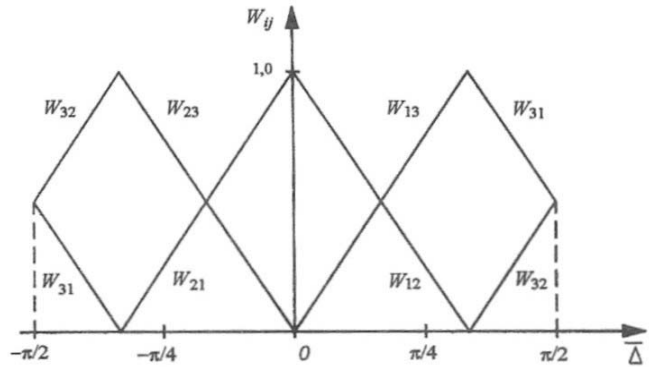

Fig. 3. Transition probabilities beyond-Neptune objects from the zone i to zone j, depending on the *ctg $\Delta'(\delta) = \overline{\Delta}$*.

The various options for the three-body problem when some special conditions [3] have predicted the parameters of orbits of hypothetical major planets, as well as the resonance zones associated with these planets, where on cosmogonic intervals of time hypothetical minor bodies can settle down. Existence of resonant objects between orbits of planets-giants also has been predicted in [15], and in [11] were calculated areas of existence resonant beyond-Neptune objects and evolution of their orbital elements is investigated. The summary table of resonant zones of works [2, 11, 15] looks like:

Table 1

| Planet | resonance | interval $a$, a.u. | | Number of the discovered objects Among Centaurs and the "Scattered-Disk" objects | Among of the Kuiper belt objects |
|---|---|---|---|---|---|
| Saturn | 2:1 | 14,826 | 15,681 | – | 2 |
| Saturn | 5:2 | 17,181 | 17,858 | – | 4 |
| Saturn | 3:1 | 19,845 | 21,207 | – | 9 |
| Saturn | 4:1 | 23,662 | 24,518 | – | 7 |
| Uranus | 5:3 | 27,265 | 27,412 | – | 1 |
| Neptune | 5:4 | 34,624 | 35,165 | 6 | 0 |
| Neptune | 4:3 | 36,062 | 36,800 | 10 | 0 |
| Neptune | 3:2 | 38,499 | 40,371 | 225 | 1 |
| Uranus | 3:1 | 39,280 | 40,686 | 191 | 1 |
| Neptune | 5:3 | 41,901 | 42,577 | 49 | 2 |
| Neptune | 7:4 | 43,374 | 43,894 | 140 | 0 |
| Neptune | 2:1 | 46,482 | 48,836 | 76 | 0 |
| «Planet 1» | 1:2 | 47,185 | 47,416 | 5 | 0 |
| Uranus | 4:1 | 47,761 | 49,432 | 20 | 1 |
| Neptune | 5:2 | 54,669 | 55,430 | – | 8 |
| Uranus | 5:1 | 56,263 | 58,123 | – | 8 |
| «Planet 1» | 2:3 | 57,207 | 57,394 | – | 3 |
| Neptune | 3:1 | 61,004 | 64,023 | – | 14 |
| Neptune | 4:1 | 73,726 | 77,459 | – | 4 |
| «Planet 1» | 4:3 | 90,852 | 91,149 | – | 1 |
| «Planet 2» | 3:4 | 123,740 | 124,222 | – | 1 |
| Neptune | 2:1 | 42,599 | 44,48 | 433 | 0 |

Here «Planet 1» and «Planet 2» - hypothetical major planets, which also includes the associated zone of proportionality (the width of zones equal to the order of $\sqrt{\mu}$) and number of objects falling in these zones. Leased line corresponds to the case of secular perturbations (3). Among the «scattered disk» objects and centaurs gets more than 26.3% in the specified area, and among the Kuiper belt objects - 79%, taking into account (3). The total bandwidth of these zones covers 17.65% of the area as 125 ÷ 5. a.u. for the centaurs and scattered disk objects, and 32.7% from the area and 30 ÷ 50. a.u. Kuiper belt objects.

**The principles in the observable distributions**

Array of found out centaurs and transneptunian objects allows to investigate significant laws in distributions of considered objects for their some orbital characteristics. With this purpose authors, according to [14] as of May, 2009 were constructed the distribution of centaurs and transneptunian objects for semi-major axis (Fig. 4), eccentricity and inclination angles of the orbits, and compared these data with theoretical results.

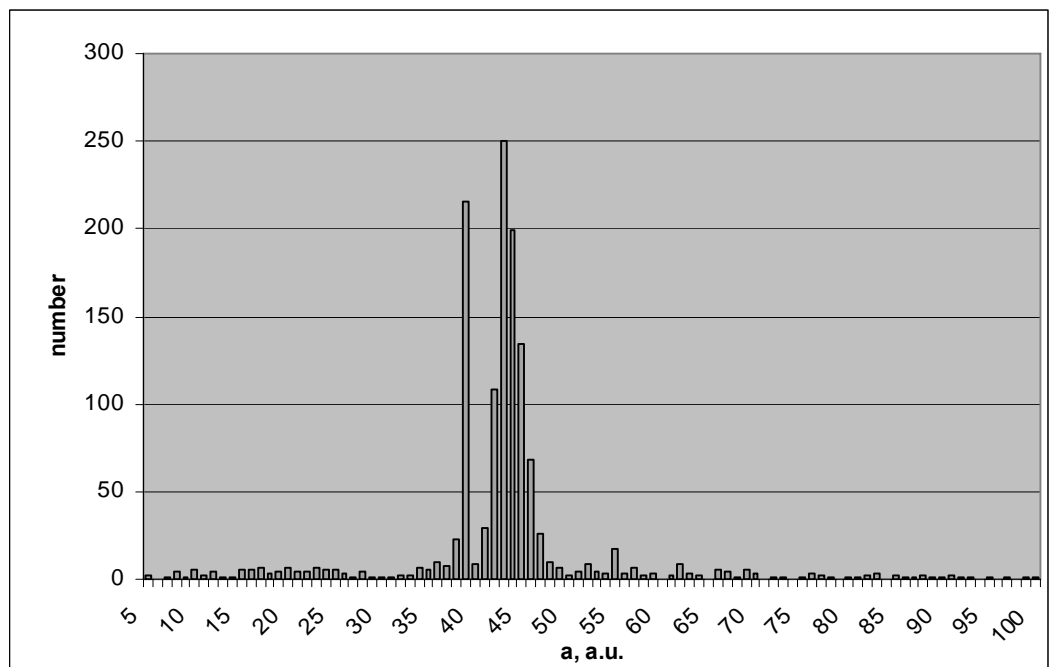

Fig. 4. The distribution of centaurs and beyond-Neptune objects for semi-major axis.

There are two obviously expressed maxima in the distributions of the Kuiper belt objects for the semi-major axes. The values of the semi-major axes, corresponding to the observable maxima of distributions 39-40, 43-44 a. u., which are within the accuracy of calculation, coincide with the predicted values in [11,15] in the range of 38.49 -40.37, 43.74-43.89 a. u. The eccentricities and inclinations orbits of beyond-Neptune objects located near the maximum to meet to value of semi-major axis of 43.5. a. u. are minor ($e \sim 0.0$-$0.17$, $i \sim 0^0$-$10^0$), that is consistent with the results of [3, 15]. For maximum charge semi-major axis at 39.8. a.u., the values of eccentricity predominantly located in the range of 0.15-0.30, while values of orbital inclination - from $0^0$ to $20^0$. From (3) implies that a correct account of secular perturbations of Uranus, Saturn and Jupiter leads to a correlation of the observed maximum of distribution (with $a$ = 43.5 a.u.) with a 2:1 resonance with Neptune. For the centaurs and the objects «scattered disk» are also clearly expressed maxima in the distribution of semi-major axis orbits, corresponding to a resonance 4:1, 5:2 with Saturn and 3:1, 5:2 with Neptune. The eccentricity of centaurs and «scattered disk» objects are predominantly in the range of $0.2 \leq e \leq 0.7$, and inclinations of their orbits are less than $30^0$.

It was found (according whith data for 2002, 2006 and 2009), that the accuracy of orbital parameters and the increase in the number of detected objects essentially does not influence a kind resulted above distributions. Coefficient of correlation between values of the same orbital parameters of objects according to 2006 and make 2009 represent 0.99, while for the corresponding series of distributions of values - from 0.91 to 0.99, depending on volume of sample. When you change the step of constructing a diagram of semi-major axis of 5 to 1 a. u. for the centaurs and «scattered disk» objects and from 1 to 0.1 a. u. for Kuiper belt objects, trends persist. Reduction of a step of digitization allows to reveal small-scale details of distribution.

With other things being equal, considered statistical distributions are influenced with selective effects (for example, farther and less bright objects it are more difficult for finding out). For smoothing selective effects has been lead wavelet - analysis of distributions which confirms invariance of a configuration of maxima of distributions (Fig. 5).

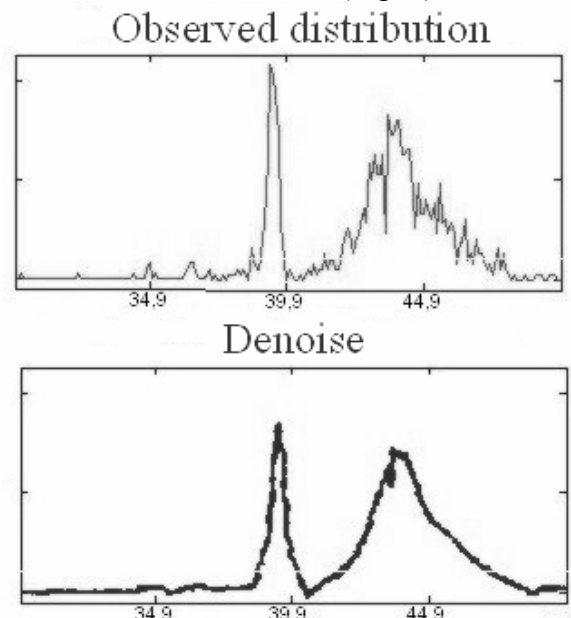

Fig. 5. Smoothing of selective effects (noise) in the distribution of Kuiper belt using wavelet.

**Conclusion**

Maxima in the distributions of semi-major axis of Kuiper belt and «irregular bodies» - centaurs and «scattered disk» objects correlate with orbital resonance zones corresponding to the maximum amplitudes of the resonance effects. Elimination of a divergence of a maximum in observable distribution of objects of Kuiper belt of semi-major axis with the theoretical value that meet the exact orbital resonance, perhaps, given the choice subdefinite integral constants, through the proper accounting of the secular perturbations of giant planets. Significant laws of distributions of investigated objects on orbital parameters are kept irrespective of size of a population of these bodies and refine of orbital parameters. In the neighborhood of the observed maxima ($a_0$) in the distributions of semi-major axis orbits, excluding selective effects, in full accordance with theoretical predictions, there is an asymmetry of distributions - with $a > a_0$, objects larger than in the case of $a < a_0$. The greater the number of investigated objects in the present work (Centaurs and "Scattered Disk" objects, the Kuiper belt objects) move on the resonant orbits and were predicted earlier.

This work has been supported by grant RFFR № 06-02-16795a.